\documentclass{mn2e}
\usepackage{psfig,mncite}

\def\figdir{./}

\def\ha{H$\alpha$}
\def\msol{M$_\odot$}
\def\teff{T$_{\textrm{\small eff}}$}

\begin{document}

\title[Spectroscopic variability of Kelu-1]{Time resolved spectroscopy of 
the variable brown dwarf Kelu-1\thanks{Based on observations obtained at 
the 
European Southern Observatory, Paranal, Chile (ESO Programme 68.C-0381)}}

\author[F.J. Clarke, C.G. Tinney and S.T. Hodgkin]{F.J. Clarke$^{1,2}$,
  C.G. Tinney$^3$ and S.T. Hodgkin$^2$\\
$^1$European Southern Observatory, Alonso de Cordova 3107, Casilla 19001, Santiago 19, Chile\\
$^2$Institute of Astronomy, Madingley Road, Cambridge CB3 0HA, UK.\\ 
$^3$Anglo-Australian Observatory, PO Box 296, Epping, NSW 2121, Australia\\
email: fclarke@eso.org, cgt@aaoepp.aao.gov.au, sth@ast.cam.ac.uk} 

\maketitle

\begin{abstract}
We report the results of observations designed to investigate the
spectroscopic signatures of dust clouds on the L2 brown dwarf
Kelu-1. Time resolved medium resolution spectra show no significant
evidence of variability in the dust sensitive TiO, CrH and FeH
bandheads on the timescale of 1--24 hours. We do however report
periodic variability in the psuedo-equivelent width of \ha\ consistent
with the 1.8 hour rotation period previously reported for this object
\cite{clarke02a}. Near-contemporaneous $I$-band photometry shows evidence
for non-periodic variability at the level of 2\%.
\end{abstract} 

\begin{keywords} 
techniques: spectroscopic, photometric --- stars: low mass, brown dwarfs
\end{keywords} 

\section{Introduction}

The brown dwarf Kelu-1 \cite{ruiz97} is a L2 dwarf in the
classification scheme of \scite{kirkpatrick99}. With an effective
temperature of $\sim$1900K, refractory molecules such as TiO, CrH and
FeH dominate the observed spectrum and play an important role in
atmospheric physics. \scite{clarke02a} (hereinafter CTC) have
discovered periodic variability in photometric observations centred on
the complex of molecular bandheads at $\sim$8600\AA. They provide
several possible explanations for the variability, including: 1) dust
cloud imhomogeneities modulating the surface brightness as Kelu-1
rotates (with an implied period of 1.8 hours), 2) A close substellar
binary inducing ellipsoidal variability (implying an orbital period of
3.6 hours). In this paper, we report the results of a search for
spectroscopic variability of Kelu-1 which answers this question.

Several spectroscopic searches have previously been made for
variability in L and T dwarfs. \scite{nakajima00} found evidence for
possible variability in the near IR water lines of the T dwarf
SDSS1624+00, and \scite{kirkpatrick01} detected changes in the
$\sim$8700\AA\ CrH feature of the L8 dwarf Gl 584C. With very similar
aims to this study, \scite{bailerjones02} has carried out near IR
spectroscopic monitoring of the variable L1.5 brown dwarf 2MASSW
J1145572+231730 (2M1145). Photometric observations of 2M1145 show
variability, but no evidence for periodicities, which
Bailer-Jones \& Mundt (2001a,b) claim as evidence for surface
features evolving on the timescale of the rotation
period. Spectroscopic observations spanning 54 hours do not show any
evidence for variability. Bailer-Jones places an upper limit of
10--15\% on the covering fraction of clear holes in a dusty
photosphere (similar to our model 3 in \S\ref{sec:modelingdust}).

In this paper we present the results of combined optical photometric
and spectroscopic observations of Kelu-1 designed to test the causes
of variability proposed by \scite{clarke02a}. Section 2 describes the
data acquisition and reductions. Analysis of the resulting spectra and
lightcurves are presented in section 3, and in section 4 we develop
toy models to investigate our observations.

\section{Observations and data reduction}

\subsection{Spectroscopy}

\begin{figure}
\centerline{%
\psfig{file=\figdir/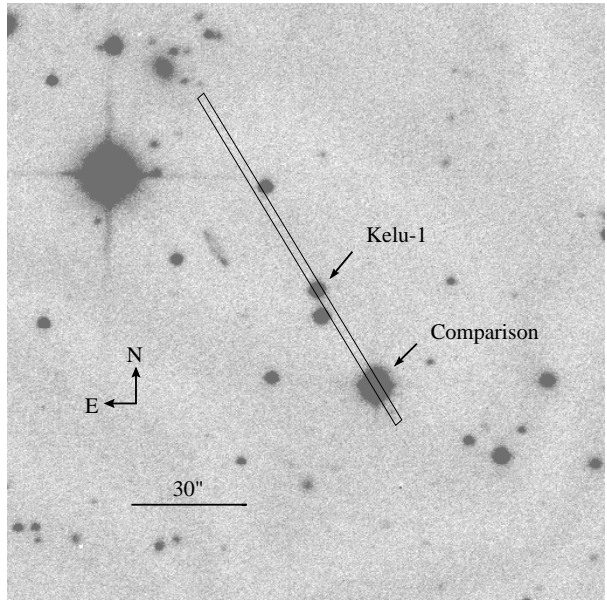,width=8.cm}}
\caption{Slit position on the sky of the FORS observations showing the
position of Kelu-1 and the comparison star.}
\label{fig:slitposition}
\end{figure}

Kelu-1 was observed on two consecutive nights (2002 February 13 and 14 UT)
with the FORS2 instrument on VLT UT4 (Yepun). Both nights were photometric,
with subarcsecond seeing. A log of observations is given in
table~\ref{tab:vltobs}. The 300I grism and 1\arcsec slit were used,
providing a spectral resolution of $\sim$13\AA\ over the range
6300--11500\AA. The OG570 order blocking filter was also used to remove
second order light. The slit was aligned such that Kelu-1 and a brighter
comparison star 25 arcsec to the south-west could be observed
simultaneously. Figure~\ref{fig:slitposition} shows the alignment of the
slit, and identifies Kelu-1 and the comparison star. The two background
stars seen near the slit in figure~\ref{fig:slitposition} are several
arcseconds from the slit, and there is no detectable scattered light from
them in our FORS2 spectra.

Each night's observations consisted of a two hour sequence of 480s
integrations (for a total of 13 spectra per night). This gave us
complete phase coverage of the 1.8 hour rotational period suggested by
CTC, and provided enough signal to noise ($\sim$70--80 per pixel;
$\sim$160--180 per resolution element) to detect small changes between
consecutive spectra. In addition to the science exposures, the ESO
calibration plan provided lamp flats and observations of the
spectrophotometric standard CD-32 9927 \cite{hamuy94}. Observations of
the standard were taken about 1 hour after the science observations on
night 2. HeAr arclamp exposures for wavelength calibration were also
taken at the end of the 2nd night.

\begin{figure}
\centerline{%
\psfig{file=\figdir/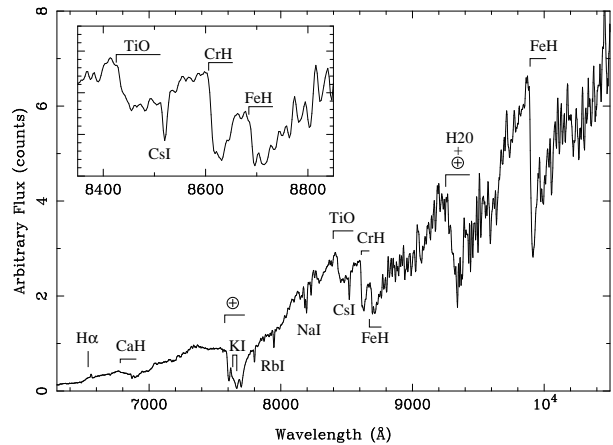,width=8.cm}}
\caption{Spectrum of Kelu-1 with important atmospheric features
denoted. The insert panel shows a close up of the molecular complex at
$\sim$8600\AA}
\label{fig:kelu1avespec}
\end{figure}

Observations were reduced with IRAF\footnote{The Image Reduction and
Analysis Facility (IRAF) is distributed by NOAO, which is operated by AURA,
Inc., under contract to the NSF.}. After subtracting a bias frame, and
correcting pixel-pixel variations and fringing with a normalized lamp flat,
the spectra were extracted with APALL task using the optimal extraction
method of \scite{horne86}. The spectra were wavelength calibrated via the
HeAr arc spectra. To keep the spectra as ``raw'' as possible, we did not
interpolate them onto a linear scale. The night sky lines show that the
wavelength solution is stable during each night, but changes by $\sim$4\AA\
between the nights. The spectrophotometric standard was then used to
correct the illumination function of the
instrument. Figure~\ref{fig:kelu1avespec} shows the averaged spectrum of
Kelu-1 from the 1st night.

\begin{table}
\centerline{%
\begin{tabular}{cccc}
\hline
Date & UT time& Seeing & Sky conditions \\
\hline
\hline
2002 Feb 13 & 05:10--07:20 & 0.75-0.9\arcsec & Photometric \\
2002 Feb 14 & 05:22--07:15 & 0.75-1.0\arcsec & Photometric \\
\hline
\end{tabular}}
\caption{Log of FORS2 spectroscopic observations carried out on VLT UT4.}
\label{tab:vltobs}
\end{table}

\subsection{Photometry}

$I$-band photometry of Kelu-1 was obtained in service mode with the
EMMI instrument at the NTT on the nights of 2002 Feburary 17 and 19
UT. ESO filter \#610 was used, which is slightly redder than the
standard Cousins $I_c$ filter. For the rest of this paper, we have
used the ESO \#610 filter profile when discussing the ``$I$''
filter. The CCD chip was windowed to give a
3.2\arcmin$\times$3.2\arcmin\ field of view. To further reduce readout
overheads we employed 2$\times$2 pixel binning, giving a pixel scale
of 0.54\arcsec/pixel. The observations consisted of consecutive 120
second exposures over two hours, with around forty frames obtained
each night.

Data were reduced in the standard fashion with IRAF, subtracting a
bias frame and dividing by a normalized dome flat. Aperture photometry
was then performed on the target and a selection of comparison stars
with the APPHOT package within IRAF. Differential lightcurves
(target$-$comparison) were then constructed in the manner described by
\scite{clarke02b}. Briefly, we produce an mean comparison star from an
ensemble of non-variable stars in the field, and it is this ``mean''
that we refer to as the comparison star in \S\ref{sec:kelu1spec_Ibandphot}.

\section{Results}

\subsection{$I$-band photometry}
\label{sec:kelu1spec_Ibandphot}

\begin{figure*}
\centerline{
\psfig{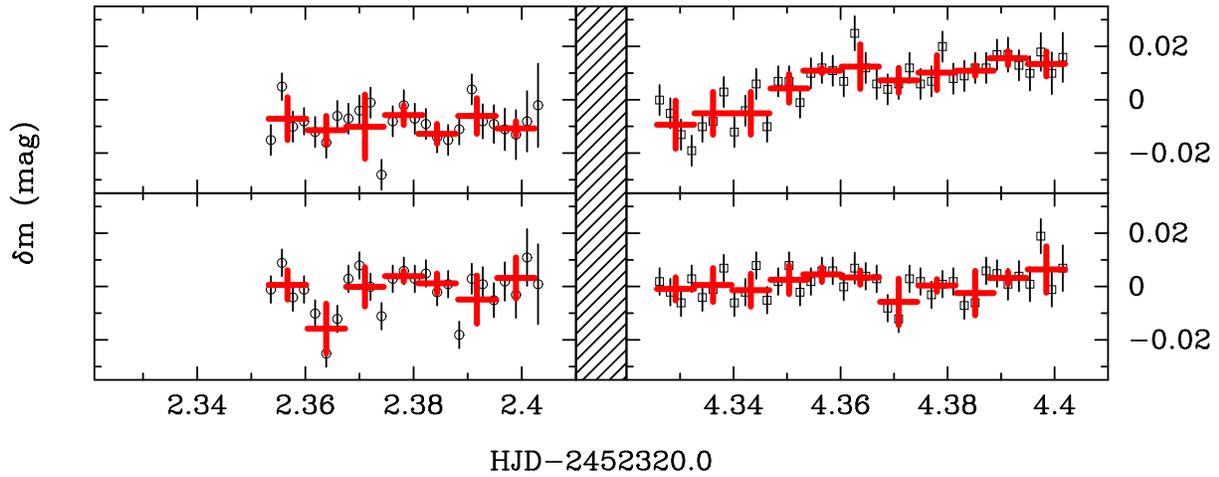}}
\caption{$I$-band lightcurve of Kelu-1 on 2002 February 17 and 19
UT. The upper panel shows Kelu-1 minus comparison star, and the lower
panel shows comparison minus check star. Each lightcurve has been mean
subtracted (using data from both nights). The thick bars show the
photometry averaged into 10 minute bins.}
\label{fig:kelu1spec_Ibandlc}
\end{figure*}

Figure~\ref{fig:kelu1spec_Ibandlc}\ shows the differential $I$-band
lightcurve of Kelu-1 minus comparison star (upper panel) and comparison
minus check star (lower panel). Table~\ref{tab:kelu1spec_photstat} gives
the statistics of the lightcurves.  The rise of 0.02 mag in the Kelu-1
lightcurve from 4.34 to 4.36 ($\sim$30 minutes) is significant at the
4-sigma level. The comparison minus check lightcurve remains constant to
0.002 mag during this time. This rise represents a dimming of Kelu-1.

\begin{table*}
\centering
\begin{tabular}{ccccccc}
\hline
Lightcurve &\multicolumn{2}{c}{Mean} & \multicolumn{2}{c}{Standard
 Dev} & \multicolumn{2}{c}{Average Dev} \\
  & night 1 & night 2 & night 1 & night 2 & night 1 & night 2 \\
\hline
\hline
Kelu-1 $-$ Comp & -0.0082 & 0.0117 & 0.0069 & 0.0054 & 0.0049& 0.0042 \\
Check $-$ Comp & -0.0013 & 0.0012 & 0.0078 & 0.0065 & 0.0057 & 0.0045\\
\hline
\end{tabular}
\caption{Statistics for Kelu-1 NTT differential photometry. The mean
for each night is mean magnitude of that night's data relative to the
mean magnitude of the whole dataset. Standard deviation is defined as
$\sqrt{\frac{1}{N-1}\sum\left({x_i-\bar{x}}\right)^2}$ and the average
deviation is defined as $\frac{1}{N}\sum\vert x_i-\bar{x}\vert$.}
\label{tab:kelu1spec_photstat}
\end{table*}

Periodogram analysis of the data does not show evidence for any
significant periods, but the analysis is heavily compromised by the
window function of the data.  They are however, inconsistent with the
1.8 hour (and 3.6 hour) period found in the 2000 photometry. As
discussed later (\S\ref{sec:ha}), there is evidence from the
spectroscopy to support the interpretation of the 1.8 hour period as
the rotation period. The $I$-band and R6 filter used by CTC
($\lambda_{\small cen}=8580$\AA, $\Delta\lambda=410$\AA) should react
similarly to inhomogeneities in the atmosphere (see
Figure~\ref{fig:models}\ and \S\ref{sec:modelingdust}), so we must
assume that the atmosphere of Kelu-1 has evolved significantly between
March 2000 and February 2002. It may be that in March 2000, Kelu-1 had
a significant long lived ($>$25 rotations) surface feature, which has
since dissipated. In February 2002, if stable features were present,
they were smaller and affected the lightcurve at less than the 0.5\%
level. The 0.02 mag dimming may be due to more rapid evolution of a
surface feature, as has been claimed for other brown dwarfs
\cite{bailerjones01,bailerjones01err}.

\subsection{Dust senstitive spectroscopic features}
\label{sec:dustfeatures}

At the temperature of Kelu-1's photosphere (\teff\,$\simeq$1900K) gas
phase molecules begin to condense into solids (i.e. dust). The removal
of species like Ti, V, Fe and Cr from the gas phase into solids
results in the weakening of TiO, VO, FeH and CrH bands. If Kelu-1's
photosphere has significant spatial dust imhomogeneities, we may
expect to see the signature in these lines (see
\S\ref{sec:modelingdust}). The complex of lines around 8700\AA\ (TiO,
CrH and FeH) is particularly interesting, as these lines dominate the
photometric band in which CTC detected variability.

\begin{figure*}
\centerline{%
\psfig{file=\figdir/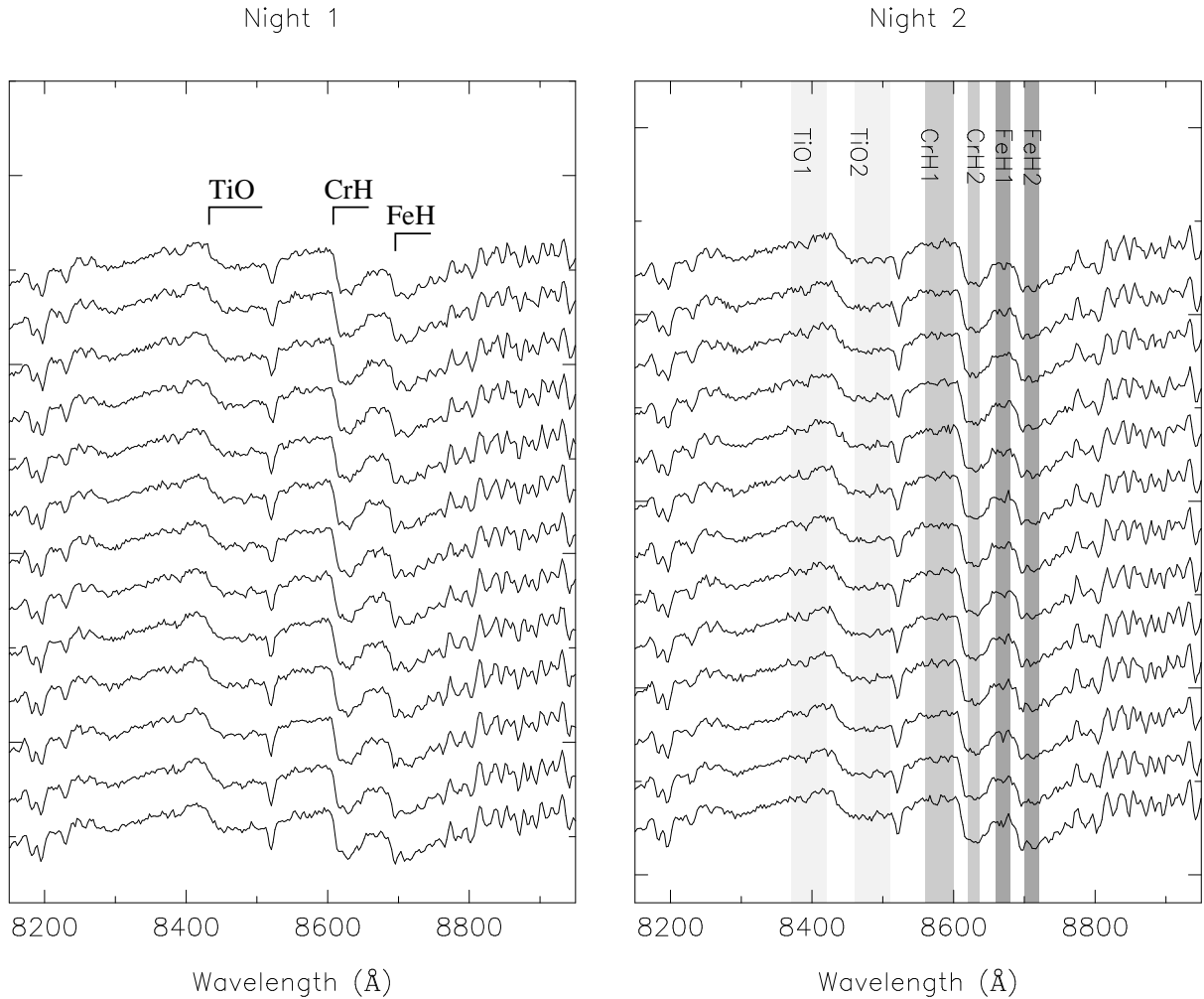,width=16.cm}}
\caption{Time series spectra of the molecular lines of TiO, CrH and
FeH at 8400--8800\AA. Time increases downwards, but the plotted gap
does not represent the actual time between observations. In the right
panel (night 2), we have marked the bands used to measure the strength
of the absorption features (Table~\ref{tab:dustbands})}
\label{fig:dustlines}
\end{figure*}

\begin{figure*} 
\centerline{%
\psfig{file=\figdir/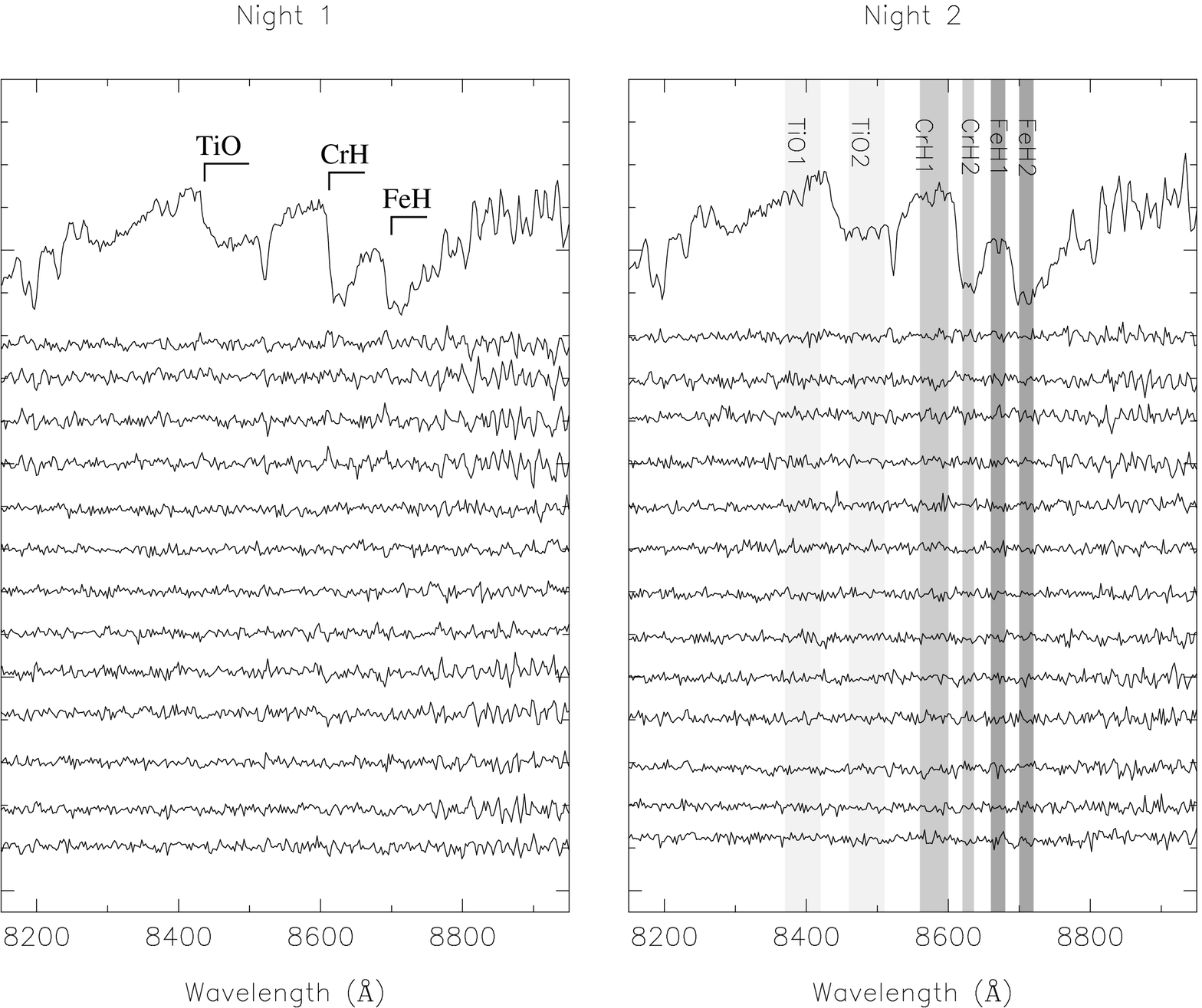,width=16.cm}
}
\caption{Difference spectra of the molecular lines of TiO, CrH and FeH
at 8400--8800\AA. Each spectrum has had the average of all the night's
spectra (upper most spectrum) subtracted. The increased noise redward
of 8800\AA\ is due to a combination of telluric variability and lower
S:N in this region ($\sim$75\% of the SN at 8600\AA).}
\label{fig:dustlines_diff}
\end{figure*}

Figure~\ref{fig:dustlines} shows the time series spectra of the
8200-8900\AA\ region, and does not show any obvious evidence for
variability. To make any small differences more apparent to the eye,
we have plotted the difference specta of this region in
figure~\ref{fig:dustlines_diff} An average spectrum has been
constructed for each night, and this has been subtracted from each
individual spectrum. In both plots, each spectrum has been offset for
clarity, with time increasing from the top of the diagram. The offsets
between consecutive spectra are constant however, and do not represent
the actual time between observations. There does not appear to be any
evidence for variability of the molecular bands either between or
during each night.

\begin{figure}
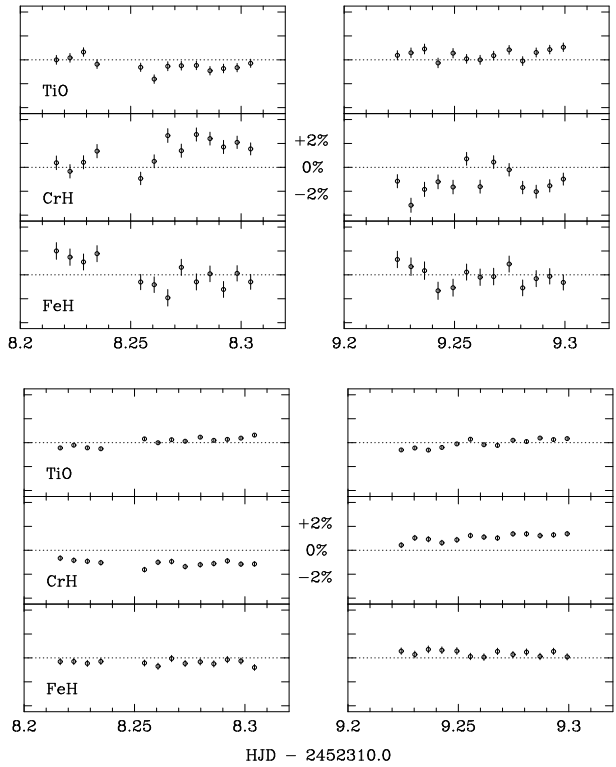

\centerline{%
\psfig{file=\figdir/kelu1bands.ps,angle=270,width=8.cm}}
\vspace{0.5cm}
\centerline{
\psfig{file=\figdir/compbands.ps,angle=270,width=8.cm}}
\caption{Changes in the molecular band strengths during the observations
for Kelu-1 (upper panels) and the comparison star (lower panels). The
definition of band strength of given in the text.}
\label{fig:dustbands}
\end{figure}

To test for subtle changes in the absorption depth, we have measured
the flux ratio between narrow bands stradling the molecular
bandheads. These bands are shown in righthand panel of
figure~\ref{fig:dustlines} and described in
table~\ref{tab:dustbands}. The flux through each band was summed, and
the molecular band index is defined as the ratio of the fluxes,
i.e. TiO index = TiO1/TiO2. To check for instrumental or atmospheric
effects, we also measured band indices for the comparison star,
although no molecular absorption is seen in this
object. Figure~\ref{fig:dustbands} shows the TiO, CrH and FeH indicies
during the observations for Kelu-1 (upper panels) and the comparison
star (lower panels). Errors are carried through from the error
estimates produced by APALL. The band indices for all these molecules
are constant to within $\pm$2\%. The most significant indication of
variability is in CrH between the nights. This is significant at the
3$\sigma$ level, but the fact that similar effects are seen in the
comparison star leads us to believe we are not observing effects
intrinsic to Kelu-1. The 1$\sigma$ upper limits on variability
(peak-to-peak) of the molecular band indices are; TiO$<$1.4\%,
CrH$<$4\%, FeH$<$2.2\%.

\begin{table}
\centerline{
\begin{tabular}{ccc}
\hline
Band & $\lambda_{\textrm{\small cen}}$ & $\Delta\lambda$ \\
\hline
\hline
TiO1 & 8395 & 50 \\
TiO2 & 8485 & 50 \\
CrH1 & 8580 & 40 \\
CrH2 & 8628 & 16 \\
FeH1 & 8670 & 20 \\
FeH2 & 8710 & 20 \\
\hline
\end{tabular}}
\caption{Bands used to measure the depth of molecular absorption features.}
\label{tab:dustbands}
\end{table}

\subsection{\ha\ Variability}
\label{sec:ha}

\begin{figure}
\centerline{%
\psfig{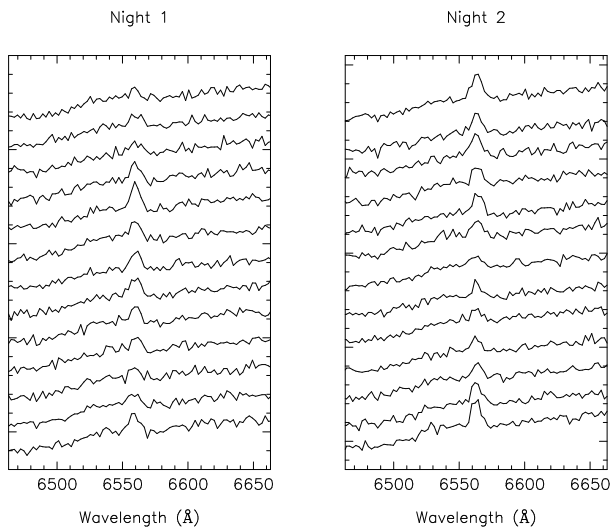}}
\caption{Close ups of the \protect\ha\ line. The first night is shown on
the left, and the second night on the right. The spectra have been offset
for clarity, with the earliest spectra in each night top of each
panel.}
\label{fig:haline}
\end{figure}

The \ha\ data show evidence for significant variability on even a
cursory examination. Figure~\ref{fig:haline} shows a close up the \ha\
line from all the spectra. Each spectrum has been offset for clarity,
in the same fashion as figure~\ref{fig:dustlines}.

\begin{figure}
\centerline{%
\psfig{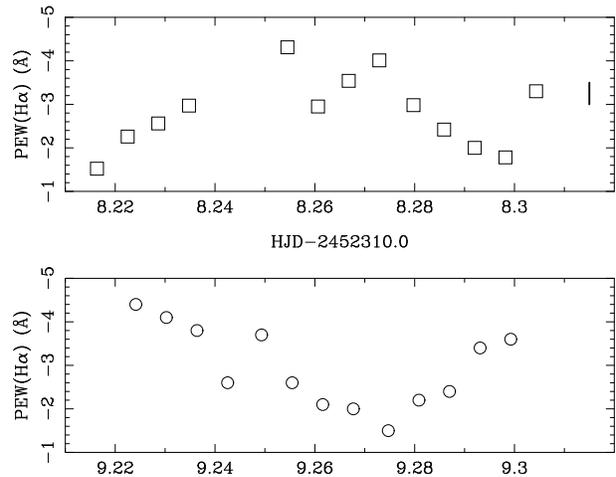}}
\caption{PEW(\protect\ha) during the FORS observations. The upper panel
shows night 1, and the lower panel night 2. Typical errors on the PEW(\ha)
are 0.5\AA, as shown by the vertical line in upper panel.}
\label{fig:haewtime}
\end{figure}

\begin{figure}
\centerline{%
\psfig{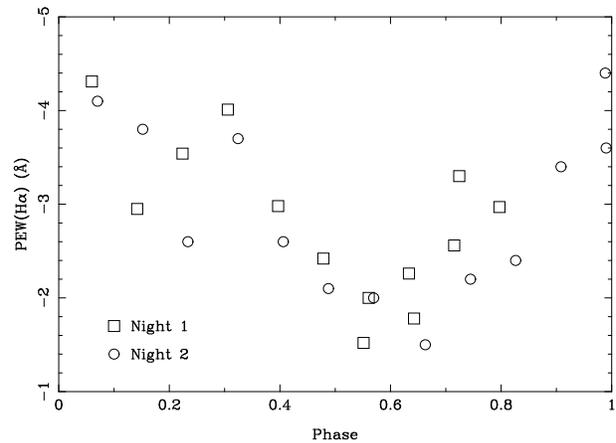}}
\caption{PEW(\ha) folded on the period of 1.8 hours reported by CTC. This
period is consistent with a periodogram analysis of the PEW(\ha) data
itself.}
\label{fig:haewphase}
\end{figure}

We have measured the psuedo equivelent width (PEW) of the \ha\ line in
all the spectra with the splot task in IRAF. The results are plotted
in figure~\ref{fig:haewtime} as PEW(\ha) as a function of time. It is
clear from this plot that the PEW(\ha) changes with time. Periodogram
analysis reveals a period in the range 1.5-2.5 hours -- consistent
with the 1.8 hour period photometric period reported by CTC. The broad
peak in the periodogram is due to the limited time coverage on each
night. Figure~\ref{fig:haewphase} shows the PEW(\ha) folded on a
period of 1.8 hours. The central wavelength of the \ha\ line does not
change significantly throughout the observations.

The mean level of PEW(\ha) is $\sim$3.0\AA\ (in emission), which is
slightly larger than, although comparable to previously measured
values (e.g. 1.5--2.0\,\AA; \pcite{basri01},\pcite{kirkpatrick99}). We
have also measured the flux in the \ha\ line with SPLOT. The mean \ha\
line flux during our observations was $0.8\pm0.4\times10^{-16}$
erg/s/cm$^2$, comparable with the value reported by \scite{ruiz97}. To
check the flux calibration of our spectroscopy, we have calculated the
flux through the $I$ bandpass. This is $\sim$75\% higher than expected
(given Kelu-1's $I$ band magnitude of 16.8, \pcite{ruiz97}), but the
discrepancy is probably due to overestimate of flux towards the red
end of the spectrum, where the standard star is very faint. We
therefore assume a conservative error of 50\% in our line flux
(included above).

\subsection{Radial velocity changes}

\begin{figure}
\centerline{%
	\psfig{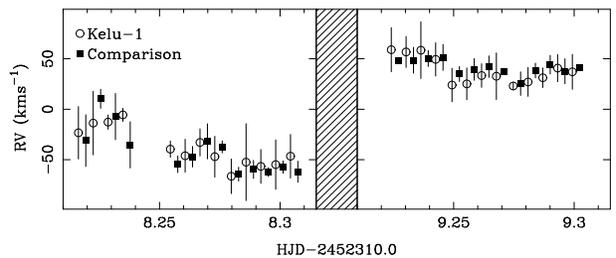}
}
\caption{Radial velocity measurements of Kelu-1 (empty circles) and the
comparison star (filled squares). The times of the comparison star points
have been slightly offset for clarity. Each meaurement is the average of
several lines, and the error bars are given by the standard deviation.}
\label{fig:velocities}
\end{figure}

To look for changes in Kelu-1's radial velocity, we have measured the
wavelength of the strong neutral atomic lines of RbI (7800, 7948\AA)
and CsI (8521\AA). These lines lie in relatively ``clean'' areas of
the spectrum and, unlike the close doublets of KI and NaI, allow
simple single profile fitting. We used the splot task in IRAF to do
this. To check for instrumental drifts, we measured the central
wavelength of several lines in the comparison star. For each line, we
convert the measured wavelength to a velocity relative to the mean
wavelength of the line in all spectra. All the lines are then averaged
to give a radial velocity for each
spectrum. Figure~\ref{fig:velocities} shows the radial velocity curves
for Kelu-1 (upper panel) and the comparison star (lower panel). The
same trend is seen in both stars, indicating the observed velocity
changes are instrumental effects. This is most likely caused by drifts
in the centering of the stars on the slit, which was typically wider
than the FWHM. The scatter on the velocity measurements of Kelu-1 is
$\pm$10\,kms$^{-1}$. This provides an upper limit on radial velocity
variations, and rules out a close companion with $M\sin
i>$10M$_{\textrm{\small jup}}$\footnote{Assuming a mass of 0.065\msol\
for Kelu-1}.  We can therefore reject the binary hypothesis for
Kelu-1's variability proposed by CTC. Note also that the lack of a 1.8
hour period in the $I$-band photometry also rules out a companion as
the source of variability, as a companion would produce consistent and
repeatable variability.

\section{Discussion}

\subsection{Limits on the inhomogeniety of Kelu-1's atmosphere}
\label{sec:modelingdust}

\scite{allard01} provide theoretical spectra for models relating to
the two limiting cases of dust formation. The DUSTY model assumes that
dust forms and remains suspended evenly throughout the photosphere,
analogous to small grains supported in thermal
updrafts. Alternatively, in the COND model, dust particles form but
immediately settle below the photosphere - a good model for grains
which are too large to be supported by convective updrafts. In
reality, the detailed physics of dust formation in substellar
atmospheres is much more complex, being a coupled system involving
chemisty, atmospheric dynamics, radiative transfer and cloud
formation. In particular, gravitational settling of dust grains and
thermal structure in the atmosphere will lead to a vertically
stratified distribution of dust. Several groups (\pcite{ackerman01},
\pcite{cooper02}, \pcite{tsuji02}, \pcite{marley02}) have made more
detailed attempts to model dust formation in substellar
atmospheres. All models to date, however, are static, and do not treat
any horizontal inhomogeneities in atmospheric structure.

To make a crude estimate of the effects a ``patchy'' photosphere would have
on the emergent spectrum, we can combine the DUSTY and COND models.  The
overall spectrum of Kelu-1 is best fit by a DUSTY model at 1900K
\cite{baraffe98}, so we assume that inhomogeneities are best modeled by
COND ``holes'' in an otherwise DUSTY photosphere. By definition, holes will
allow us to see into deeper, hotter regions of the photosphere. With this
in mind, we have constructed several models;

\newcounter{model}
\begin{list}%
{Model--\arabic{model}}{\usecounter{model}\setlength{\leftmargin}{0.25cm}\setlength{\rightmargin}{\leftmargin}}
\item 2100K COND holes representing patches of effcient dust settling
which allow us to see deeper into the atmosphere.
\vspace{0.1cm}
\item 2100K DUSTY holes representing patches which allow us to see a
deeper level of the atmosphere, but where dust is still significant.
\vspace{0.1cm}
\item 1900K COND holes representing patches of efficent dust settling at
  the same temperature as the rest of the photosphere.
\end{list}

\noindent
In all cases, the 1900K DUSTY spectrum represents the majority of the
photosphere. To model the effects of imhomogeneties in the atmosphere, we
construct a linear combination of the two spectra as;

\begin{equation}
f(\lambda)_{\textrm{mix}} = (1.0-F)f(\lambda)_{\mbox{phot}} + Ff(\lambda)_{\mbox{hole}}
\label{eq:mixspectra}
\end{equation}

\noindent
where $f(\lambda)_{\mbox{phot}}$ and $f(\lambda)_{\mbox{hole}}$ are
the model spectra representing the normal and ``hole'' regions of the
atmosphere, and $F$ is the covering fraction of holes. We therefore
generate a ``mixed'' spectrum, for which we can measure molecular band
indices in exactly the same way as for the real data
(\S\ref{sec:dustfeatures}). We also measure the photometric bands $I$
(ESO \#610) and R6 (as used by CTC; $\lambda_{\small cen}=8580$\AA,
$\Delta\lambda=410$\AA).

\begin{figure*}
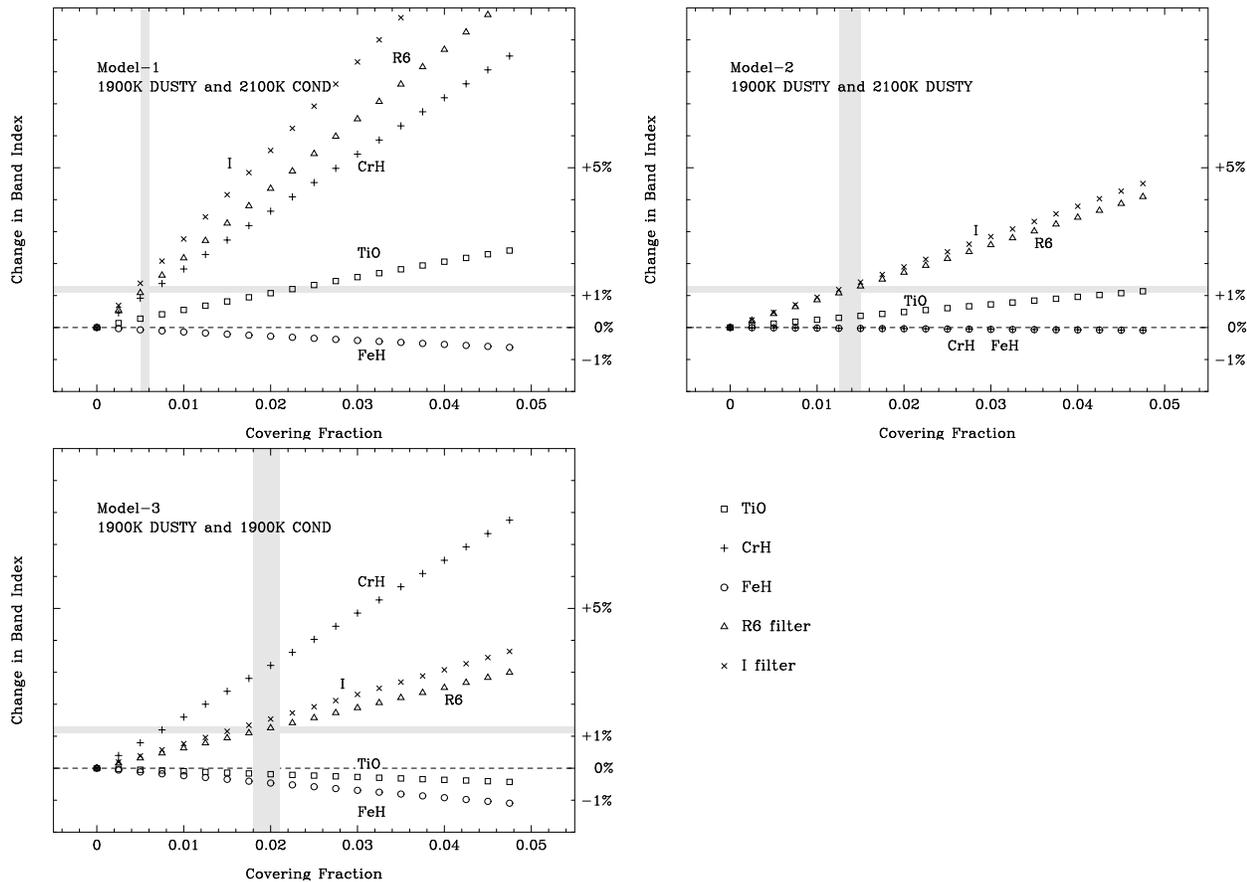

\begin{tabular}{cc}
\psfig{file=\figdir/1900d2100c.ps,angle=270,width=8.cm}&
\psfig{file=\figdir/1900d2100d.ps,angle=270,width=8.cm}\\
\psfig{file=\figdir/1900d1900c.ps,angle=270,width=8.cm}&
\psfig{file=\figdir/modelkey.ps,angle=270,width=8.cm}\\
\end{tabular}
\caption{Changes in band strength with covering fraction for all three
models. Model--1 upper left, Model--2 upper right and Model--3 lower
left. The shaded areas in each plot represent the amplitude of the R6
variability discovered by CTC (1.2$\pm$0.1\%), and the corresponding
surface covering fraction. A covering fraction of 0 represents a
photosphere with no ``holes''.}
\label{fig:models}
\end{figure*}

Figure~\ref{fig:models} shows the effects of changing the ``hole''
covering fraction for all three models. The band indices are
normalised to the purely dusty case (no holes, $F=0$), so the points
give the change in band indices we would observe if the photosphere
changed from having no holes, to having a given covering fraction. The
effect of small changes in $F$ is roughly linear, so the plots are
also valid for a changes between two non-zero covering fractions.

The horizontal shaded area shows the R6 photometric amplitude reported by
CTC (1.2$\pm$0.1\%). The most striking feature of these plots is the
incredibly small changes in covering fraction (vertical shaded area)
required to reproduce the observed photometric variability; less than 2\%
``holes'' in all three models. Comparing with the molecular band indices,
our upper limits on variability (TiO$<$1.4\%; CrH$<$4\%; FeH$<$2.2\%) are
entirely consistent with these covering fractions for all models. It should
be noted however, that these limits represent the \textit{difference} in
covering fraction between one side of the brown dwarf and the other, the
atmosphere may be far more inhomogenous on small scales.

Comparing the models reveals interesting differences between them. The
CrH index appears to be much more sensitive to dust properities
(models 1 \& 3) rather than temperature (model 2). The TiO index
offers a potential descriminant between models 1 \& 3, becoming
stronger with covering fraction in model 1, and weaker in model
3. None of the molecular bands are strongly effected in model 2, which
depends purely on temperature. The fact that the scatter in our
measurement of CrH is larger than TiO and FeH is more in line with
models 1 and 3 than model 2. However, the most probable explanation of
differences between the models is that our toy models are too
primitive to throw any light onto these subtle effects. It is likely a
full 3-dimensional model of a brown dwarfs atmosphere will be required
to interpret the results of future spectroscopic studies.

\subsection{The chromosphere of Kelu-1}

The \ha\ emission we see is probably caused by ionised hydrogen gas in
Kelu-1's chromosphere, analogous to other low mass stars
(e.g. \pcite{gizis00}). The energy required to heat the gas may come
from magnetic heating \cite{mohanty02}, or wave heating
\cite{yelle00}, although calculations of the energy released from both
are problematic. Combined with magnetic fields above the surface of
Kelu-1, this could lead to inhomogenous patches of hot gas above
Kelu-1 and produce the variable \ha\ emission we observe. Similar
periodic \ha\ variability has been observed from other
stars. \scite{fernandez98} show the variable \ha\ emission from
several weaklined T-Tauri stars is correlated with photometric
variability, which they interpret as chromospheric magnetic loops
corotating above cool magnetic spots in the photosphere. Similar
anchored magnetic loops could exist above Kelu-1, leading to \ha\
variability with the same (rotation) period as photospheric
variability.

The presence of magnetic fields and a chromosphere do not necessarily
contradict previous arguments against magnetic spots in the
photospheres of ultra cool dwarfs
\cite{bailerjones01,gelino02,mohanty02,bailerjones02}. Densities in
the region of the photosphere are much higher, and the atmosphere
there is much more neutral. This corresponds to very low magnetic
Reynolds numbers \cite{gelino02}, or high electrical resistivities
\cite{mohanty02}. In the region of the photosphere, the magnetic field
is decoupled from the atmospheric fluid. A magnetic field may be
generated deeper in the atmosphere, where the temperature and
ionisation fraction rise, and the magnetic field is well coupled to
the matter.

The \ha\ line flux we measured in \S\ref{sec:ha} corresponds to an
\ha\ luminosity of $3.7\pm1.8\times10^{24}$ erg/s/cm$^{2}$ at Kelu-1's
distance of 19.6\,pc \cite{dahn02}. In the standard measure of
chromospheric activity, Kelu-1 has
log(L(\ha)/L\raisebox{-0.5ex}[0ex][0ex]{\scriptsize bol}) of
$-5.35\pm0.5$, where our error estimates are quite conservative
(\S\ref{sec:ha}). This value is in line with what we would expect for
a L2 dwarf from the work of \scite{gizis00} and \scite{mohanty02}.

\section{Conclusions}

We have presented here the first high signal-to-noise phase resolved
optical spectroscopy of a brown dwarf with a known rotation period. We
have detected a rotational modulation of the \ha\ line, consistent
with the 1.8 hour rotation period reported by
\scite{clarke02a}. Radial velocity measurements reject the possibility
that photometric variability is caused by a close companion to
Kelu-1. Photometry shows that the atmosphere of Kelu-1 has
significantly evolved between March 2000 and February 2002, when no
periodic signal can be detected at the level of 0.5\%. Kelu-1 does
however show evidence for a 2\% dimming over $\sim$30 minutes. We have
also placed upper limits on the scale of possible surface features
inducing inhomogenous dust formation across the surface, and made a
primitive attempt to investigate the effects of different atmosphere
models. On the balance of observations, it seems that Kelu-1 is a
single brown dwarf, with a typical level of chromospheric activity,
and a relatively homogenous, although evolving, atmosphere.

\section{Acknowledgements}

We would like to thank the staff of the VLT and NTT for their help in
making these observations, especially for rearranging at short notice
the NTT queue to allow quasi-simultaneous observations. We also thank
the referee, Coryn Bailer-Jones, for his timely response and useful
comments. FJC acknowledges the support of a PPARC studentship award
during the course of this research.

\bibliography{browndwarfs,thesis} 
\bibliographystyle{mn}

\end{document}